\documentclass[aps,prl,twocolumn]{revtex4}

\usepackage{graphicx}
\usepackage{color}
\usepackage{bm}
\usepackage{amsmath, amsthm, amssymb,mathrsfs}
\usepackage{csquotes}
\usepackage[colorlinks=true, citecolor=blue,allcolors=blue]{hyperref}

\begin{document}

\title{Strong-Field Ionization Phenomena Revealed by Quantum Trajectories}

\author{Taylor Moon$^{1,2}$, Klaus Bartschat$^3$, and Nicolas Douguet$^{1,4}$}

\affiliation{$^1$Department of Physics, Kennesaw State University, Marietta, 30060, USA\\
$^{2}$Department of Geosciences, The University of Montana, Missoula, MT 59808, USA\\
$^3$Department of Physics and Astronomy, Drake University, Des Moines, Iowa 50311, USA\\
$^4$Department of Physics, University of Central Florida, Orlando, FL, 32816, USA}

\date{\today}
\email{nicolas.douguet@ucf.edu}


\begin{abstract}
We investigate the photoionization dynamics of atoms subjected to intense, ultrashort laser pulses through the use of quantum trajectories. This method provides a unique and consistent framework for examining electron dynamics within a time-dependent potential barrier. Our findings demonstrate that quantum trajectories offer additional insights into several key aspects of strong-field ionization, including the transition between ionization regimes, non-adiabatic effects under the barrier, the impact of the shape of the electronic potential, and the efficiency of over-the-barrier ionization.
\end{abstract}

\maketitle

Recent progress in generating bright and ultra\-short laser pulses \cite{Paul01,Li20,Maroju2020,Zhang17}
has profoundly transformed our approach to study light-matter interactions. In particular, these advances have opened the way
to observe and manipulate the electron dynamics in atoms and molecules at their intrinsic temporal scale \cite{Stewart23,Isinger17,Sainadh2019}.
In spite of these remarkable strides, the theoretical modeling of strong-field phenomena, which is pivotal for the advancement of attosecond science, remains challenging.
One major difficulty arises from the complexity in describing and comprehending the electron dynamics under the combined influence of the Coulomb interaction 
between charged particles and an intense electromagnetic field \cite{Popruzhenko_2014,Faisal_1973,Reiss80,AGOSTINI2012117,Smirnova08,Torlina12}.

\begin{figure}[b]
\vspace{-5.0truemm}
\includegraphics[width=8.cm]{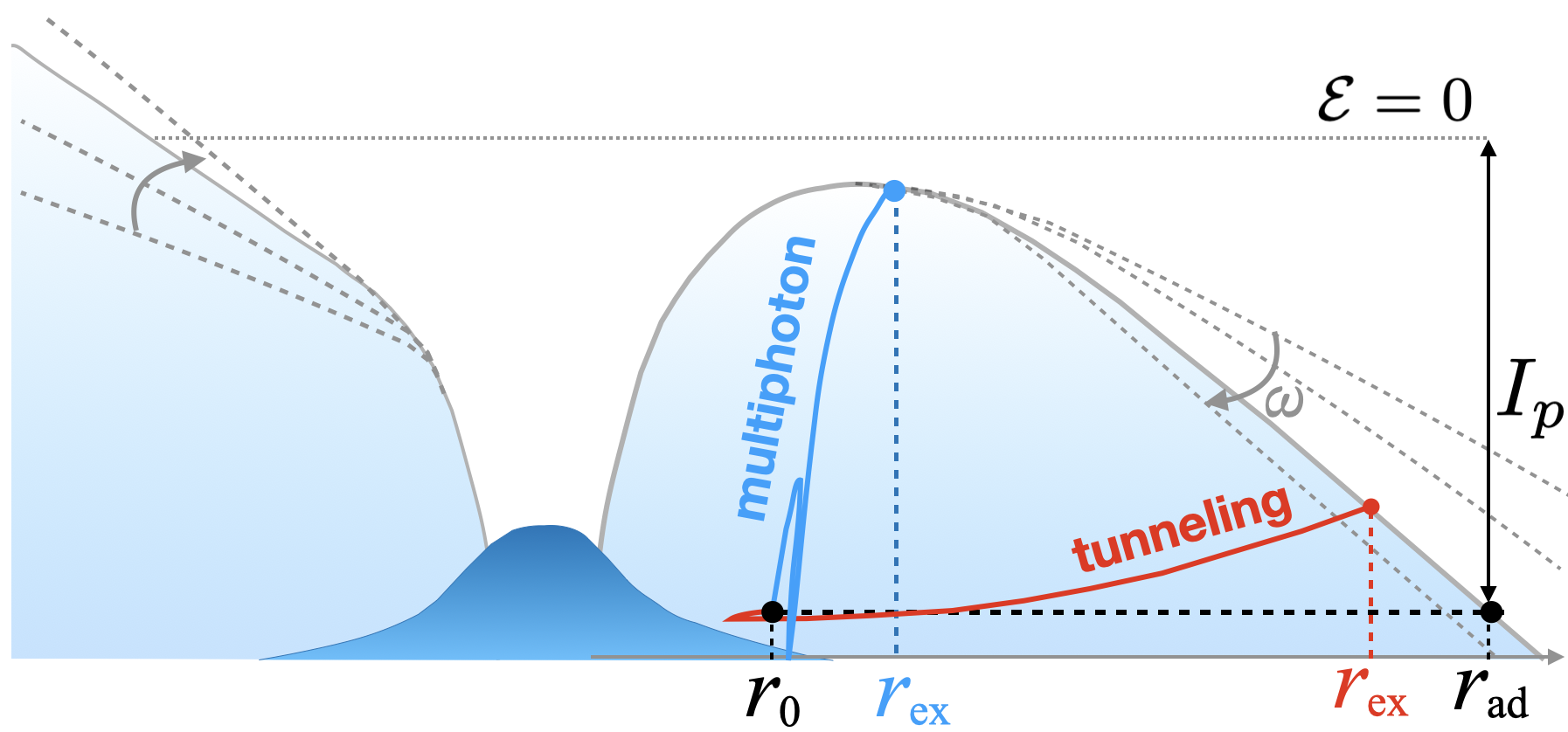}
\caption{Schematic illustration of multi\-photon (vertical channel) and tunneling (horizontal channel) ionization. See text for the definition of the various parameters. The trajectories shown are the results of numerical calculations with the formalism presented in this paper.}
\label{fig:1}
\end{figure}

When an atom is subjected to the oscillating electric field of a laser beam, its electronic wave function is perturbed
by the time-varying potential barrier. If sufficient energy is transferred to the atom, this process 
eventually leads to the emission of one or many electrons. One usually distinguishes two main types of photoionization mechanisms, 
as illustrated in Fig.~\ref{fig:1}.
In the {\it vertical} channel, referred to as multi\-photon ionization (MPI) \cite{Delone2000,Ivanov05},
the barrier oscillates rapidly compared to the characteristic response time of the system, 
such that the electron experiences periodic ``heating" due to the swift changes of the potential.
Hence, the electron accumulates an average energy during each pulse cycle, gradually drifting away from the ionic core until it becomes free to escape. 
This regime of ionization is chaotic \cite{Ivanov05} and dominates for weak, high-frequency fields.
Conversely, in the {\it horizontal} channel, known as tunneling ionization (TI), the wave function has sufficient time to adjust to the gradual changes of the potential barrier, 
allowing its tail to ``leak" via tunneling through the quasi-static barrier. 
In general, however, the MPI and TI regimes coexist, i.e., an electron can  always gain energy non\-adiabatically inside a moving barrier.

The Keldysh parameter, $\gamma_{\rm \footnotesize\textsc{k}}\equiv\omega\sqrt{2I_p}/E_0$~\cite{Keldysh}, 
provides a quantitative understanding of the dominant photo\-ionization mechanism by comparing
the period of the oscillating electric field $T$ to the characteristic response time $\tau$ of the system. By taking $\tau$ as the tunneling time of the electron from
the inner to the outer classical turning points, a transparent expression for the Keldysh parameter is obtained \cite{Topcu12,Keldysh,Delone_1998,Ivanov05}, 
depending only on the light's angular frequency
$\omega$, the maximum field strength $E_0$, and the ionization potential $I_p$ of the atom. The MPI picture is privileged when $\gamma_{\rm \footnotesize\textsc{k}}\gg1$,
whereas the TI mechanism dominates for $\gamma_{\rm \footnotesize\textsc{k}}\le1$.

While the Keldysh parameter provides a simple analytical formula revealing the dominant photoionization mechanism, it also has 
severe limitations \cite{Reiss2010,Popruzhenko_2014,Topcu12}. For instance, it disregards the nature of the binding potential and the role of excited states \cite{Trombetta89,Mishima02}. 
Moreover, $\gamma_{\rm \footnotesize\textsc{k}}$ only considers the pulse at its maximum strength and does not account for the details of the pulse, such as its duration, 
envelope, and carrier-envelope phase (CEP), despite their known critical roles in strong fields \cite{Chetty22,Karamatskou13}. It also wrongly assumes that the tunnelling wave packet 
always starts from the classically allowed region \cite{Douguet18},
while the possibility for over-the-barrier ionization (OBI) \cite{KRAINOV95,Schuricke11} (i.e., when the top of the barrier gets below the initial electronic state energy) is neglected. 
Despite theoretical improvements \cite{Arbo08,Popruzhenko08,Smirnova08,Torlina12,Topcu12,Wang19,Zheltikov16,Mishima02}, there remains a critical need 
for quantitative approaches unraveling the electron dynamics in strong fields \cite{Popruzhenko_2014}.

On the other hand, the dynamics of an electron exiting a potential barrier becomes rapidly classical \cite{Heldt23,Xie22,Ni16}, thus enabling an 
alternative description via trajectory-based methods \cite{Camus17,Salieres01,Yang16,Tan21}. These approaches, extensively used in strong-field physics, 
offer an appealing avenue to bridge quantum and classical ideas \cite{Xie22,Ni16}.
Notably, they form the cornerstone of the three-step model \cite{PhysRevLett.68.3535,corkum1993,Corkum07} 
and can be integrated in more sophisticated methodologies, e.g., analytic tunneling rates with imaginary time methods \cite{Perelomov66}
or Wigner phase-space distributions \cite{Hack21} to improve the treatment of strong-field phenomena.

In this Letter, we utilize de Broglie-Bohm quantum trajectories \cite{duerr2009bohmian,Botheron10,Benseny14,Jooya15,Zimmermann16} 
to establish a comprehensive and consistent framework that encompasses both the short-range region, where quantum effects dominate, 
and the large-distance region, where electron trajectories behave classically. This methodology serves as a powerful tool for extracting intricate information embedded in the time-dependent wave function. 
Notably, we introduce a quantum-trajectory adiabatic parameter that extends the Keldysh parameter, revealing detailed insights into the dynamics 
of an electron under a potential barrier. Additionally, this framework makes it possible to quantify the efficiency of the OBI mechanism.

Our approach to compute quantum trajectories starts with solving 
the time-dependent Schr\"{o}dinger equation (TDSE) \cite{Douguet16}, 
$i\partial_t|\Psi\rangle=\hat{H}(t)|\Psi\rangle$, in the single-active electron (SAE) picture. 
[Unless stated otherwise, atomic units (a.u.) are used throughout this manuscript.]
In the Coulomb gauge and dipole approximation,
$\hat{H}(t)=\hat{H}_0+\hat{H}_{\rm int}(t)$, where $\hat{H}_0=\hat{T}+\hat{V}$ is the field-free hamiltonian, $\hat{T}$ and $\hat{V}$ are the kinetic and potential energies, 
respectively, and \hbox{$\hat{H}_{\rm int}=-\bm p\cdot {\bm A}(t)$} is the light-atom interaction in the velocity gauge. 
The vector potential, ${\bm A}(t)=-A_0F(t)\sin{(\omega t})\,\hat{{\bm x}}$, has a period $T$, is 
linearly polarized along the $x$ axis, has an amplitude $A_0$, and an envelope \hbox{$F(t)=\cos^4(\pi t/NT)$} for $N$ cycles with $t\in[-NT/2,NT/2]$. 
The electric field, ${\bm E}(t)=-d{\bm A}(t)/dt$, reaches its maximum strength $E_0$, corresponding to an intensity $I=E_0^2/2$, near the envelope maximum at $t=0$.

In the Bohmian formalism \cite{duerr2009bohmian}, the velocity field at a position $\bm r$ and time $t$ is given 
by $\bm v(\bm r,t)=\bm j(\bm r,t)/\rho(\bm r,t)$, where $\bm j(\bm r,t)={\rm \Re}\left[\Psi^*(\bm r,t)\left(-i\nabla+{\bm A}(t)\right)\Psi(\bm r,t)\right]$ 
is the probability flux density, ${\rm \Re}[z]$ denotes the real part of $z$, and $\rho(\bm r,t)=|\Psi(\bm r,t)|^2$ is the probability density. The velocity field
is used to propagate the trajectories starting from different initial position $\bm r_0$ in the initial state. These trajectories describe the flow of the probability density, 
and they make it possible to reconstruct the wave function and estimate the energy of the system at any given time \cite{Song_2017}. Furthermore, they allow us to retrieve
any quantum observables with high accuracy \cite{Xie22}.

Using the intrinsic properties of these trajectories, one can devise complementary quantities to investigate strong-field phenomena.
Specifically, we focus at first on the efficiency of the horizontal and vertical ionization channels (see Fig.~\ref{fig:1}) by computing the exit radius $r_{\rm ex}$ 
and exit time $t_{\rm ex}$
from the potential barrier for trajectories starting at different initial radii~$\bm r_0$.
These exit quantities are found from the condition $\Delta \mathcal{E}(t_{\rm ex})=0$, 
where $\Delta \mathcal{E}(t)=v^2(t)/2+Q(t)$ is 
the gauge-independent energy difference
between the energy $\mathcal{E}(t)$ of the trajectory and the barrier energy~\cite{Kobe_1987}. 
The quantum potential 
$Q=-1/2\{\rho^{-1/2}\nabla^2\rho^{1/2}\}$
is responsible for all quantum effects and the non-zero velocity at the tunneling exit \cite{Kim_2021,Ivanov24,Han19}. In the classical limit ($Q=0$),
the equations of motion reduce to the classical Hamilton-Jacobi equations.
The adiabatic radius, $r_{\rm ad}$, corresponds to the radius associated with adiabatic TI, i.e., for a trajectory 
maintaining a constant energy $\mathcal{E}=-I_p$ and exiting the barrier at the same time $t_{\rm ex}$ and in the same direction 
as the trajectory under consideration. Thus, in contrast to standard approaches, which define adiabatic tunneling at the maximum field strength, 
$r_{\rm ad}$ is a dynamical quantity that takes into account the shape of the barrier at the exit time. 
Finally, in the subsequent discussion, we exclusively focus on ionizing trajectories, which 
exhibit asymptotic freedom, i.e., $\mathcal{E}(\infty)\ge0$.

\begin{figure}[b]
\includegraphics[width=\columnwidth]{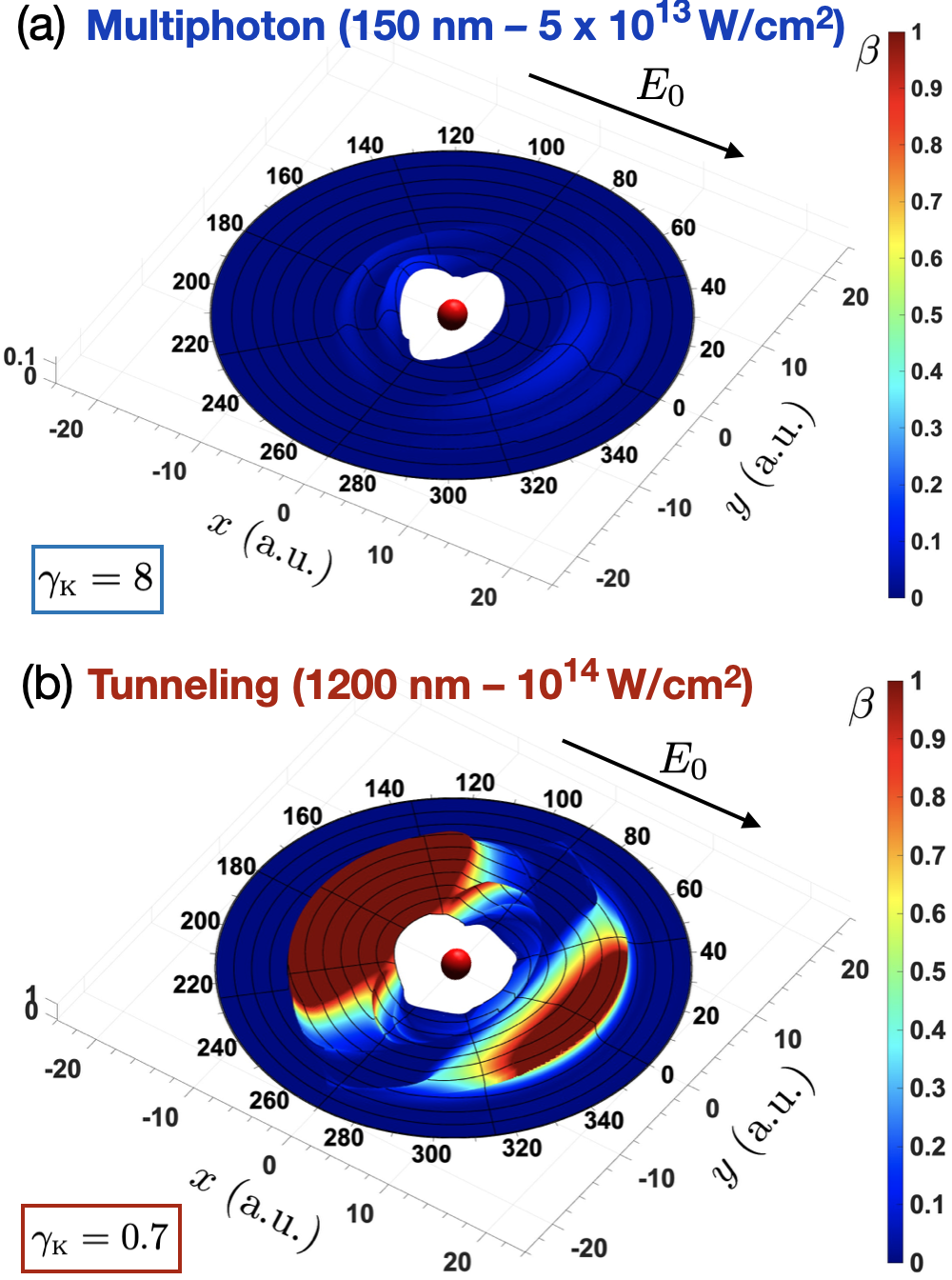}
\caption{Tunneling parameter $\beta$ from the hydrogen ground state depending on the initial position of the trajectories in a plane containing the field. 
The nucleus is schematically represented by the red sphere at the origin. (a) MPI case (2-cycle, 150~nm, 5$\times$10$^{13}\,$W/cm$^2$ pulse); 
(b) TI case (1-cycle, 1200~nm, 1$\times$10$^{14}\,$W/cm$^2$ pulse). The arrows indicate the direction of maximum electric field. 
The inner white area corresponds to the non\-ionizing region. See text for details.}
\label{fig:2}
\end{figure}

\begin{figure*}[t]
\includegraphics[width=18cm]{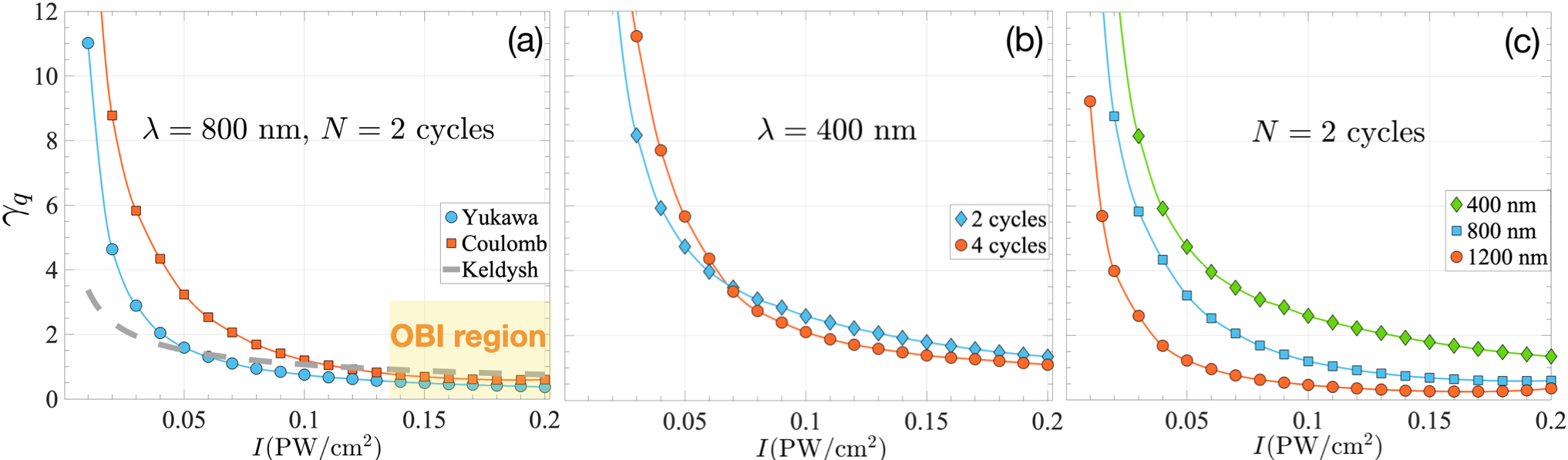}\
\caption{Value of the quantum-trajectory adiabatic parameter $\gamma_q$ as a function of the peak intensity in various situations. The pulse characteristics are indicated in the panels. 
Comparison (a) for two different potentials (Coulomb and Yukawa) and the standard Keldysh parameter ($I_p=0.5$~a.u.) (b) for different pulse lengths in H($1s$) and (c) for various wavelengths in H($1s$).}
\label{fig:3}
\end{figure*}

We now define a tunneling parameter 
\begin{equation}
\beta(\bm r_0)\equiv
\frac{r_{\rm ex}(\bm r_0) -r_0}{r_{\rm ad}(\bm r_0)-r_0},
\end{equation} 
with $0\le\beta\le1$,
to quantify 
the degree of tunneling in a trajectory starting at $\bm r_0$.
For vertical ionization, $r_{\rm ex}=r_0$ and $\beta=0$, whereas for perfect horizontal ionization,
$r_{\rm ex}=r_{\rm ad}$ and $\beta=1$. 
Perfect tunneling, $\beta=1$, is also assigned in the event that $r_{\rm ex}\ge r_{\rm ad}$ and for OBI, as detailed below. 
If a multi\-photon trajectory exits the potential barrier more than once, $r_{\rm ex}$ is taken at the first exit time. 
As for the Keldysh parameter, this definition considers the length-gauge picture for tunneling. However, 
the definition of $\beta$ is not unique and could be refined further or tailored for the study of specific phenomena.

In Fig.~\ref{fig:2}, we present two-dimensional plots of $\beta(\bm r_0)$ for photoionization of atomic hydrogen as a function of 
the initial positions ($r_0<20$~a.u.) of the trajectories in the $xy$ plane. We employ two radically different pulses, 
describing an MPI case ($\gamma_{\rm \footnotesize\textsc{k}}=8$), and a TI situation ($\gamma_{\rm \footnotesize\textsc{k}}=0.7$) 
dominated by a single ionization event. 
The observed values of $\beta$ closely align with our expectations, exhibiting small values throughout ($0\le\beta\le0.1$) 
for the vertical ionization case and a large region indicating tunneling for the long-wavelength scenario.

Remarkably, $\beta$ not only confirms these expectations but also provides insightful and detailed information regarding the ionization mechanism. 
In Fig.~\ref{fig:2}(b), it is evident that the ionization process strongly involves tunneling along the field, 
with larger ionization occurring from the region opposite to the direction of maximum field strength.
Conversely, the multi\-photon mechanism occurs orthogonal to the field, where the potential barrier is weak or non\-existent.
Therefore, this mapping provides opportunities to study the angular-resolved tunneling dynamics. 
At large distances, the mechanism naturally exhibits multi\-photon character, since the trajectories exit at the early stage of the pulse, 
with a broad barrier to cross and no time to respond to changes in the field.
Monitoring the time-dependent energy of these trajectories reveals a more chaotic behavior for multi\-photon trajectories, 
while tunneling trajectories exhibit a smooth and traceable energy variation, 
rapidly converging to that of a classical particle ($Q\to0$) after exiting the barrier. This method also enables the retrieval of the exit position and momentum of the electron.

The statistically averaged tunneling level of the ionization process, $\bar\beta={W_0}^{-1}\int_{\mathcal{D}}\beta(\bm r_0)\rho(\bm r_0,0)d^3\bm r_0$, 
is obtained by weighting $\beta(\bm r_0)$ over the initial probability density carried by each ionizing trajectory \cite{Garashchuk03}. 
Here, $\mathcal{D}$ denotes the set of ionizing trajectories and $W_0=\int_{\mathcal{D}}\rho(\bm r_0,0)d^3\bm r_0$ is the total ionization probability. To design a quantity directly comparable 
with the Keldysh parameter, we define the quantum adiabatic parameter, $\gamma_q\equiv(1-\bar\beta)/\bar\beta$, as the ratio between the 
degrees of MPI and TI. This quantity has the same limits as $\gamma_{\rm \footnotesize\textsc{k}}$, 
i.e., $\gamma_q=0$ for perfect TI, $\gamma_q\to\infty$ as the ionization regime tends to MPI, and $\gamma_q=1$ 
for equal MPI and TI efficiency ($\bar\beta=1/2$). 

Figure~\ref{fig:3}(a) compares $\gamma_q$ and $\gamma_{\rm \footnotesize\textsc{k}}$ for H($1s$) across various intensities of an 800-nm pulse. 
While both parameters approach unity at similar intensities, $\gamma_q$ shows a more abrupt transition between ionization regimes, deviating from the typical $I^{-1/2}$ law. 
For weak fields, MPI is significantly more pronounced than predicted by the Keldysh approximation. 
This discrepancy arises because $\gamma_{\rm \footnotesize\textsc{k}}$ is based on the concept of tunneling through a distorted potential, 
rather than the stepwise energy absorption characteristic of MPI \cite{Trombetta89}.
To explore the sensitivity of $\gamma_q$ to the potential shape, we also considered the short-range Yukawa potential, $V(r)=-Z_{0}e^{-r}/r$, 
which is often used to highlight tunneling effects \cite{Torlina15}. 
By setting $Z_0 = 1.908$ to reproduce the hydrogen ground-state energy ($I_p = 0.5$~a.u.), $\gamma_{\rm \footnotesize\textsc{k}}$ remains unchanged. 
However, $\gamma_q$ indicates significantly more TI for the Yukawa potential compared to the Coulomb potential in the intensity range considered. 
At low intensities, the presence of excited states in the Coulomb potential naturally enhances MPI. As the intensity increases, tunneling becomes the dominant mechanism, 
and the larger imaginary velocity in the Yukawa potential close to the nucleus provides a stronger initial ``kick” during tunneling. In addition, the response time in Yukawa is faster due to the absence of excited states.
At very high intensities, the imaginary velocity is primarily influenced by the field strength, causing $\gamma_q$ for both Yukawa and Coulomb potentials to be more similar.
Nevertheless, the Coulomb potential always induces quantum trajectories that tunnel slightly earlier than in the Yukawa potential \cite{Klaiber15}.


In Fig.~\ref{fig:3}(b), we analyze the changes of $\gamma_q$ with the pulse duration, considering a 400-nm pulse with either $N=2$ or 4 cycles. 
A subtle difference is evident between the two pulses: $\gamma_q$ for the 4-cycle pulse exhibits increased MPI at low intensity 
and more prominent tunneling at high intensity compared to the \hbox{2-cycle} pulse. This difference is explained as follows:
While both pulses reach the same field maximum at $t=0$, their satellite peaks, at $t=\pm T/2$, differ due to the different pulse envelopes.
Consequently, within the MPI regime, vertical excitation is more prominent for the 4-cycle pulse due to the higher and additional satellite peaks.
With rising intensity, the efficiency of TI, which follows an exponential law with the field strength \cite{Ammosov86}, 
increases more rapidly at the satellite peaks of the 4-cycle pulse.
Therefore, the difference in $\gamma_q$ arises mainly from the contrasting tunneling efficiency between the primary and the side peaks for each pulse. 
As tunneling at the primary peak becomes increasingly dominant with rising field strength, the two curves tend to converge at high intensity levels, 
as seen in Fig.~\ref{fig:3}(b).

Next, we computed $\gamma_q$ for a 2-cycle pulse at three different wavelengths.
While the results shown in Fig.~\ref{fig:3}(c) follow the general predictions, the curves at different wavelengths cannot be obtained by simple rescaling, 
as predicted by the Keldysh model. Even for $\gamma_{\rm \footnotesize\textsc{k}}<1$, non-adiabatic effects are still important.
The electron has a coordinate-dependent energy, which makes it tunnel out closer to the core than predicted by a static tunneling picture \cite{Klaiber15}. This effect is
further amplified by the Coulomb potential \cite{Klaiber15}, explaining the slow decrease rate of $\gamma_q$.
As the intensity increases further, ionizing trajectories start nearer to the classical turning point, with a reduced imaginary velocity and, consequently, a longer tunneling time. 
This results in a small, but noticeable inflection in $\gamma_q$, occurring at lower intensities for the $1200$~nm pulse compared to the $800$~nm pulse, due to its slightly higher ionization rate. 
This inflection is absent for the Yukawa potential in the examined intensity range, explained by its smaller ionization rate and classical turning point, as well as its larger imaginary velocity. 
A detailed analysis of the variation of $\gamma_q$ at higher intensities will be presented in a future study. Note that the inflection in $\gamma_q$ only corresponds to a phase transition.
As we delve deeper into the OBI regime, $\gamma_q$ will start decreasing again.

 \begin{figure}[t!]
\includegraphics[width=\columnwidth]{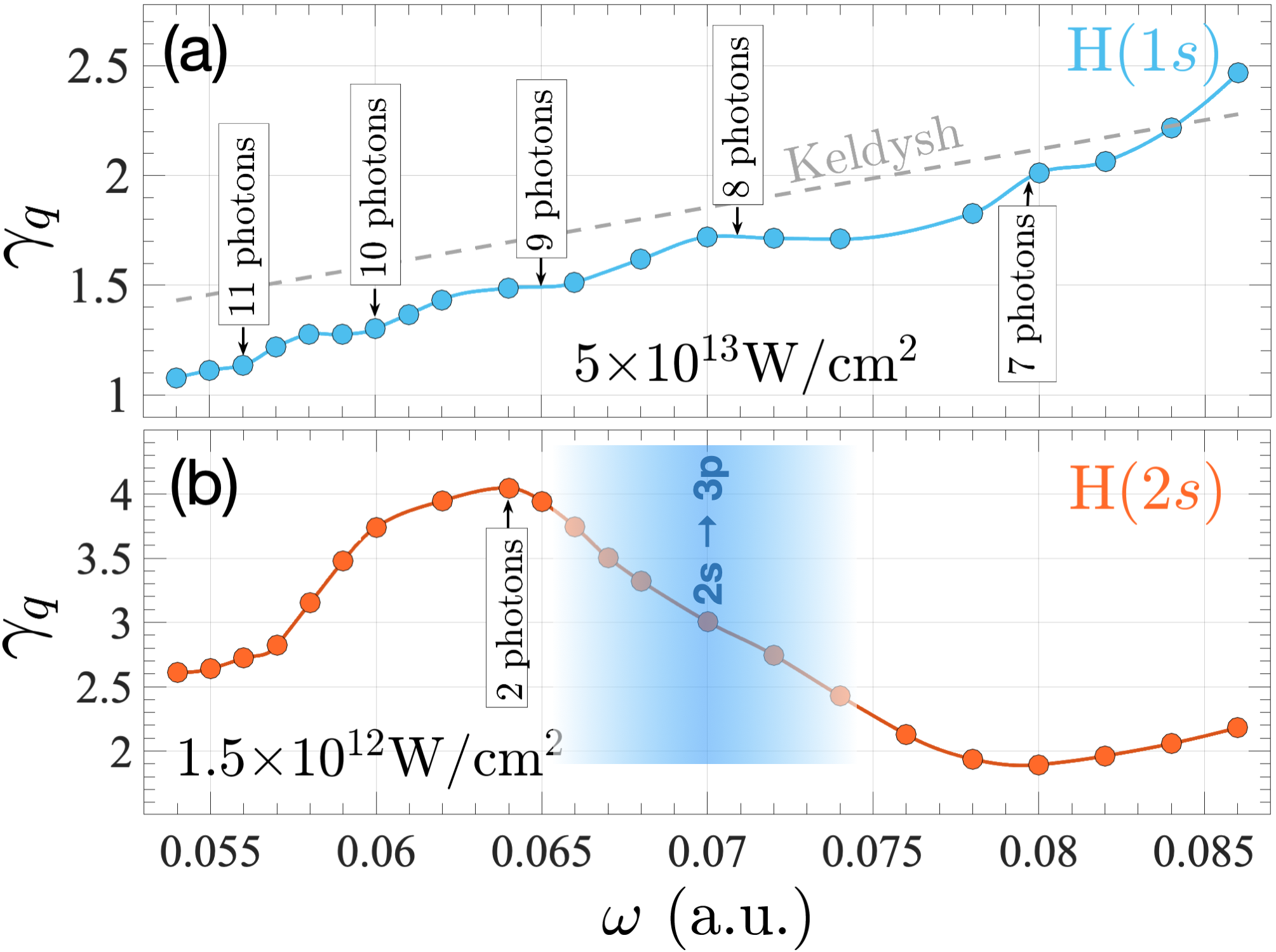}\\
\caption{$\gamma_q$ for a $2$-cycle pulse as a function of $\omega$ starting (a) from the H($1s$) ground state compared with $\gamma_{\rm \footnotesize\textsc{k}}$, 
and (b) from the H($2s$) state near the $2s\to3p$ transition energy. The frequencies for $m$-photon channel closing are indicated by arrows 
with the minimum number of photons to ionize also indicated. See text for details.}  
 \label{fig:4}
 \end{figure}

\begin{figure}[b!]
\includegraphics[width=8.5cm]{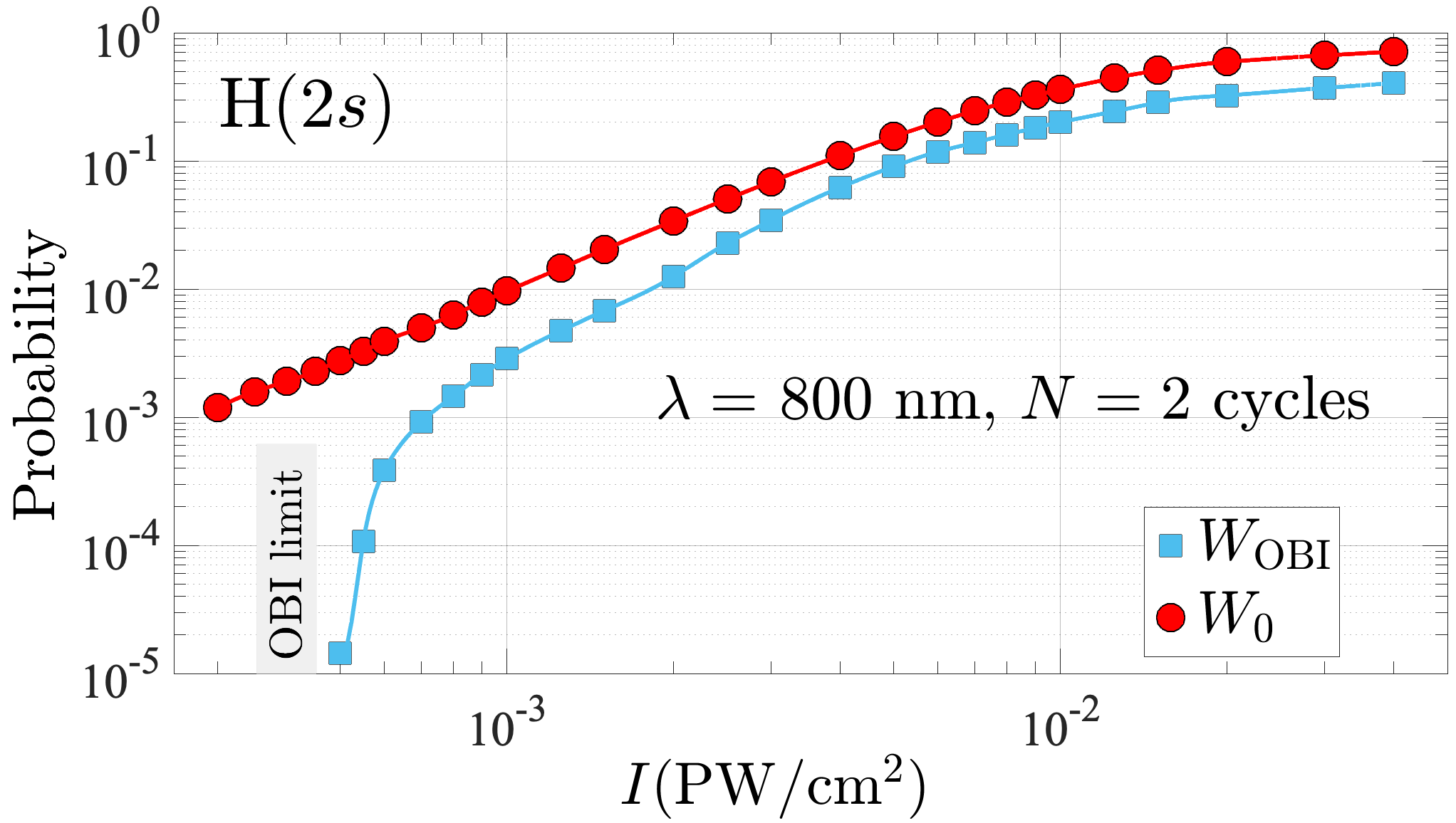}
\caption{Total ionization probability and OBI probability starting from the H($2s$) state. The standard OBI limit is indicated on the horizontal axis.}
\label{fig:5}
\end{figure}

Returning to the analysis of the frequency-dependent sensitivity of $\gamma_q$, 
Fig.~\ref{fig:4}(a) presents its variation as a function of $\omega$ for a 2-cycle pulse with $I=5\times10^{13}\,$W/cm$^2$.
While $\gamma_q$ exhibits an overall linear increase with~$\omega$, 
akin to $\gamma_{\rm \footnotesize\textsc{k}}$, 
we observe intermittent steps at which MPI ceases to progress uniformly. 
These distinctive steps correlate with channel-closing events \cite{Zimmermann17}, i.e., a change in the minimum number $m$ of 
photons needed to ionize the system. An $m$-photon channel is closed when $m\omega\le I_p +U_p$, where $U_p=E_0^2/4\omega^2$ is the ponderomotive energy, 
which effectively shifts the ionization threshold upward. 
At channel closing, the excitation of the quasi-continuum of Rydberg states is known
to suppress ionization, because the characteristic time of the electron's orbital motion is much longer than the optical cycle \cite{Topcu12}. 
As a result, $\gamma_q$ varies negligibly near channel closing.

In order to highlight the potential impact of excited states on the ionization process, we present in Fig.~\ref{fig:4}(b) 
results for two-photon ionization of atomic hydrogen in the $2s$ state with varying frequencies around the $2s \to 3p$ transition. 
Rather than showing a steady increase, a sharp drop in $\gamma_q$ occurs near resonance, indicating a preference for TI in this frequency region. 
This phenomenon is due to population transfer to the $3p$ state, which temporally stabilizes the wave-packet energy and thus favors indirect TI, confirming findings reported in \cite{Mishima02}. 
As opposed to excited Ryd\-berg states, the electron in the $3p$ orbital remains close to the nucleus, 
and the width of the barrier in this energy range is sufficiently large for efficient horizontal ionization to occur. 
With increasing intensity, the effect of resonant excitation on $\gamma_q$ should be less pronounced.

As a final application, we explore OBI via quantum trajectories, ionizing the H($2s$) state with a 800-nm pulse. 
An ionizing trajectory is classified as exhibiting OBI characteristics if either its energy $\mathcal{E}(t)$ consistently 
remains above the barrier or if it exits the barrier with $\mathcal{E}(t_{\rm ex})\le I_p$.  Using this definition, one can
compute an OBI ionization probability, denoted as $W_{\rm OBI}$, which is compared to $W_0$ in Fig.~\ref{fig:5}. $W_{\rm OBI}$ 
experiences a sharp increase above the OBI limit corresponding to the intensity where the top of the potential barrier becomes 
lower than $I_p$. With increasing intensity, $W_{\rm OBI}$ starts to rise at a similar rate as $W_0$, passing through a 
transition phase where all trajectories starting from the classically allowed region go over the barrier. Only at an intensity 
about ten times larger than the OBI limit, OBI reaches a steady regime where the ratio $W_{\rm OBI}/W_0\approx0.6$ remains nearly constant. 
We verified that the trajectories that do not exhibit OBI characteristics correlate with ionization away from the field polarization.

In conclusion, we showed that quantum trajectories represent a powerful tool to investigate strong-field phenomena 
and explore the electron dynamics beneath and above a potential barrier. 
We demonstrated that this framework can provide quantitative information to shed light on various high-intensity effects, such as indirect TI, channel closing, and exit momenta.
In particular, we reveal how the Keldysh parameter underestimates MPI at low intensities, and put in evidence specific non-adiabatic effects and the crucial role of the long-range electronic potential.
Lastly, we quantified the intensity-dependent efficiency of OBI.
This methodology provides fine details about the TI mechanism, potentially paving the way for exploring novel dynamical effects and advancing theoretical models, 
e.g., the study of recombining and rescattering trajectories in high-order harmonic generation (HHG) \cite{Ishii08,Dong20,Petrakis21,Piper22} and light-induced electron diffraction (LIED) \cite{DeGiovannini_2023}, 
the phenomenon of frustrated tunneling ionization \cite{Nubbemeyer08}, or the influence of chirality on tunnel-ionization dynamics \cite{Bloch21}.

\smallskip
This work was supported by the NSF under grants \hbox{No.~PHY-2012078 } (N.D.) and \hbox{No.~PHY-2110023} (K.B.).

\bibliography{Bohmian-2.bib}

\begin{thebibliography}{65}
\expandafter\ifx\csname natexlab\endcsname\relax\def\natexlab#1{#1}\fi
\expandafter\ifx\csname bibnamefont\endcsname\relax
  \def\bibnamefont#1{#1}\fi
\expandafter\ifx\csname bibfnamefont\endcsname\relax
  \def\bibfnamefont#1{#1}\fi
\expandafter\ifx\csname citenamefont\endcsname\relax
  \def\citenamefont#1{#1}\fi
\expandafter\ifx\csname url\endcsname\relax
  \def\url#1{\texttt{#1}}\fi
\expandafter\ifx\csname urlprefix\endcsname\relax\def\urlprefix{URL }\fi
\providecommand{\bibinfo}[2]{#2}
\providecommand{\eprint}[2][]{\url{#2}}

\bibitem[{\citenamefont{Paul et~al.}(2001)\citenamefont{Paul, Toma, Breger,
  Mullot, Augé, Balcou, Muller, and Agostini}}]{Paul01}
\bibinfo{author}{\bibfnamefont{P.~M.} \bibnamefont{Paul}},
  \bibinfo{author}{\bibfnamefont{E.~S.} \bibnamefont{Toma}},
  \bibinfo{author}{\bibfnamefont{P.}~\bibnamefont{Breger}},
  \bibinfo{author}{\bibfnamefont{G.}~\bibnamefont{Mullot}},
  \bibinfo{author}{\bibfnamefont{F.}~\bibnamefont{Augé}},
  \bibinfo{author}{\bibfnamefont{P.}~\bibnamefont{Balcou}},
  \bibinfo{author}{\bibfnamefont{H.~G.} \bibnamefont{Muller}},
  \bibnamefont{and} \bibinfo{author}{\bibfnamefont{P.}~\bibnamefont{Agostini}},
  \bibinfo{journal}{Science} \textbf{\bibinfo{volume}{292}},
  \bibinfo{pages}{1689} (\bibinfo{year}{2001}).

\bibitem[{\citenamefont{Li et~al.}(2020)\citenamefont{Li, Lu, Han, Li, Wu,
  Wang, Ghimire, and Chang}}]{Li20}
\bibinfo{author}{\bibfnamefont{J.}~\bibnamefont{Li}},
  \bibinfo{author}{\bibfnamefont{J.}~\bibnamefont{Lu}},
  \bibinfo{author}{\bibfnamefont{S.}~\bibnamefont{Han}},
  \bibinfo{author}{\bibfnamefont{J.}~\bibnamefont{Li}},
  \bibinfo{author}{\bibfnamefont{Y.}~\bibnamefont{Wu}},
  \bibinfo{author}{\bibfnamefont{H.}~\bibnamefont{Wang}},
  \bibinfo{author}{\bibfnamefont{S.}~\bibnamefont{Ghimire}}, \bibnamefont{and}
  \bibinfo{author}{\bibfnamefont{Z.}~\bibnamefont{Chang}},
  \bibinfo{journal}{Nat. Commun.} \textbf{\bibinfo{volume}{11}},
  \bibinfo{pages}{2748} (\bibinfo{year}{2020}).

\bibitem[{\citenamefont{Maroju et~al.}(2020)\citenamefont{Maroju, Grazioli,
  Di~Fraia, Moioli, Ertel, Ahmadi, Plekan, Finetti, Allaria, Giannessi
  et~al.}}]{Maroju2020}
\bibinfo{author}{\bibfnamefont{P.~K.} \bibnamefont{Maroju}},
  \bibinfo{author}{\bibfnamefont{C.}~\bibnamefont{Grazioli}},
  \bibinfo{author}{\bibfnamefont{M.}~\bibnamefont{Di~Fraia}},
  \bibinfo{author}{\bibfnamefont{M.}~\bibnamefont{Moioli}},
  \bibinfo{author}{\bibfnamefont{D.}~\bibnamefont{Ertel}},
  \bibinfo{author}{\bibfnamefont{H.}~\bibnamefont{Ahmadi}},
  \bibinfo{author}{\bibfnamefont{O.}~\bibnamefont{Plekan}},
  \bibinfo{author}{\bibfnamefont{P.}~\bibnamefont{Finetti}},
  \bibinfo{author}{\bibfnamefont{E.}~\bibnamefont{Allaria}},
  \bibinfo{author}{\bibfnamefont{L.}~\bibnamefont{Giannessi}},
  \bibnamefont{et~al.}, \bibinfo{journal}{Nature}
  \textbf{\bibinfo{volume}{578}}, \bibinfo{pages}{386} (\bibinfo{year}{2020}).

\bibitem[{\citenamefont{Zhang et~al.}(2017)\citenamefont{Zhang, Bucklew,
  Edwards, Janisch, and Liu}}]{Zhang17}
\bibinfo{author}{\bibfnamefont{C.}~\bibnamefont{Zhang}},
  \bibinfo{author}{\bibfnamefont{V.}~\bibnamefont{Bucklew}},
  \bibinfo{author}{\bibfnamefont{P.}~\bibnamefont{Edwards}},
  \bibinfo{author}{\bibfnamefont{C.}~\bibnamefont{Janisch}}, \bibnamefont{and}
  \bibinfo{author}{\bibfnamefont{Z.}~\bibnamefont{Liu}}, \bibinfo{journal}{Opt.
  Lett.} \textbf{\bibinfo{volume}{42}}, \bibinfo{pages}{502}
  (\bibinfo{year}{2017}).

\bibitem[{\citenamefont{Stewart et~al.}(2023)\citenamefont{Stewart, Hoerner,
  Debrah, Lee, Schlegel, and Li}}]{Stewart23}
\bibinfo{author}{\bibfnamefont{G.~A.} \bibnamefont{Stewart}},
  \bibinfo{author}{\bibfnamefont{P.}~\bibnamefont{Hoerner}},
  \bibinfo{author}{\bibfnamefont{D.~A.} \bibnamefont{Debrah}},
  \bibinfo{author}{\bibfnamefont{S.~K.} \bibnamefont{Lee}},
  \bibinfo{author}{\bibfnamefont{H.~B.} \bibnamefont{Schlegel}},
  \bibnamefont{and} \bibinfo{author}{\bibfnamefont{W.}~\bibnamefont{Li}},
  \bibinfo{journal}{Phys. Rev. Lett.} \textbf{\bibinfo{volume}{130}},
  \bibinfo{pages}{083202} (\bibinfo{year}{2023}).

\bibitem[{\citenamefont{Isinger et~al.}(2017)\citenamefont{Isinger, Squibb,
  Busto, Zhong, Harth, Kroon, Nandi, Arnold, Miranda, Dahlstr\"om
  et~al.}}]{Isinger17}
\bibinfo{author}{\bibfnamefont{M.}~\bibnamefont{Isinger}},
  \bibinfo{author}{\bibfnamefont{R.~J.} \bibnamefont{Squibb}},
  \bibinfo{author}{\bibfnamefont{D.}~\bibnamefont{Busto}},
  \bibinfo{author}{\bibfnamefont{S.}~\bibnamefont{Zhong}},
  \bibinfo{author}{\bibfnamefont{A.}~\bibnamefont{Harth}},
  \bibinfo{author}{\bibfnamefont{D.}~\bibnamefont{Kroon}},
  \bibinfo{author}{\bibfnamefont{S.}~\bibnamefont{Nandi}},
  \bibinfo{author}{\bibfnamefont{C.~L.} \bibnamefont{Arnold}},
  \bibinfo{author}{\bibfnamefont{M.}~\bibnamefont{Miranda}},
  \bibinfo{author}{\bibfnamefont{J.~M.} \bibnamefont{Dahlstr\"om}},
  \bibnamefont{et~al.}, \bibinfo{journal}{Science}
  \textbf{\bibinfo{volume}{358}}, \bibinfo{pages}{893} (\bibinfo{year}{2017}).

\bibitem[{\citenamefont{Sainadh et~al.}(2019)\citenamefont{Sainadh, Xu, Wang,
  Atia-Tul-Noor, Wallace, Douguet, Bray, Ivanov, Bartschat, Kheifets
  et~al.}}]{Sainadh2019}
\bibinfo{author}{\bibfnamefont{U.~S.} \bibnamefont{Sainadh}},
  \bibinfo{author}{\bibfnamefont{H.}~\bibnamefont{Xu}},
  \bibinfo{author}{\bibfnamefont{X.}~\bibnamefont{Wang}},
  \bibinfo{author}{\bibfnamefont{A.}~\bibnamefont{Atia-Tul-Noor}},
  \bibinfo{author}{\bibfnamefont{W.~C.} \bibnamefont{Wallace}},
  \bibinfo{author}{\bibfnamefont{N.}~\bibnamefont{Douguet}},
  \bibinfo{author}{\bibfnamefont{A.}~\bibnamefont{Bray}},
  \bibinfo{author}{\bibfnamefont{I.}~\bibnamefont{Ivanov}},
  \bibinfo{author}{\bibfnamefont{K.}~\bibnamefont{Bartschat}},
  \bibinfo{author}{\bibfnamefont{A.}~\bibnamefont{Kheifets}},
  \bibnamefont{et~al.}, \bibinfo{journal}{Nature}
  \textbf{\bibinfo{volume}{568}}, \bibinfo{pages}{75} (\bibinfo{year}{2019}).

\bibitem[{\citenamefont{Popruzhenko}(2014)}]{Popruzhenko_2014}
\bibinfo{author}{\bibfnamefont{S.~V.} \bibnamefont{Popruzhenko}},
  \bibinfo{journal}{J.\ Phys.\ B: At.\ Mol.\ Opt.\ Phys.}
  \textbf{\bibinfo{volume}{47}}, \bibinfo{pages}{204001}
  (\bibinfo{year}{2014}).

\bibitem[{\citenamefont{Faisal}(1973)}]{Faisal_1973}
\bibinfo{author}{\bibfnamefont{F.~H.~M.} \bibnamefont{Faisal}},
  \bibinfo{journal}{J.\ Phys.\ B: At.\ Mol.\ Opt.\ Phys.}
  \textbf{\bibinfo{volume}{6}}, \bibinfo{pages}{L89} (\bibinfo{year}{1973}).

\bibitem[{\citenamefont{Reiss}(1980)}]{Reiss80}
\bibinfo{author}{\bibfnamefont{H.~R.} \bibnamefont{Reiss}},
  \bibinfo{journal}{Phys. Rev. A} \textbf{\bibinfo{volume}{22}},
  \bibinfo{pages}{770} (\bibinfo{year}{1980}).

\bibitem[{\citenamefont{Agostini and DiMauro}(2012)}]{AGOSTINI2012117}
\bibinfo{author}{\bibfnamefont{P.}~\bibnamefont{Agostini}} \bibnamefont{and}
  \bibinfo{author}{\bibfnamefont{L.~F.} \bibnamefont{DiMauro}}, in
  \emph{\bibinfo{booktitle}{Adv.\ At.\ Mol.\ Opt.\ Phys.}}, edited by
  \bibinfo{editor}{\bibfnamefont{P.}~\bibnamefont{Berman}},
  \bibinfo{editor}{\bibfnamefont{E.}~\bibnamefont{Arimondo}}, \bibnamefont{and}
  \bibinfo{editor}{\bibfnamefont{C.}~\bibnamefont{Lin}}
  (\bibinfo{publisher}{Academic Press}, \bibinfo{year}{2012}),
  vol.~\bibinfo{volume}{61}, pp. \bibinfo{pages}{117--158}.

\bibitem[{\citenamefont{Smirnova et~al.}(2008)\citenamefont{Smirnova, Spanner,
  and Ivanov}}]{Smirnova08}
\bibinfo{author}{\bibfnamefont{O.}~\bibnamefont{Smirnova}},
  \bibinfo{author}{\bibfnamefont{M.}~\bibnamefont{Spanner}}, \bibnamefont{and}
  \bibinfo{author}{\bibfnamefont{M.}~\bibnamefont{Ivanov}},
  \bibinfo{journal}{Phys. Rev. A} \textbf{\bibinfo{volume}{77}},
  \bibinfo{pages}{033407} (\bibinfo{year}{2008}).

\bibitem[{\citenamefont{Torlina and Smirnova}(2012)}]{Torlina12}
\bibinfo{author}{\bibfnamefont{L.}~\bibnamefont{Torlina}} \bibnamefont{and}
  \bibinfo{author}{\bibfnamefont{O.}~\bibnamefont{Smirnova}},
  \bibinfo{journal}{Phys. Rev. A} \textbf{\bibinfo{volume}{86}},
  \bibinfo{pages}{043408} (\bibinfo{year}{2012}).

\bibitem[{\citenamefont{Delone and Krainov}(2000)}]{Delone2000}
\bibinfo{author}{\bibfnamefont{N.~B.} \bibnamefont{Delone}} \bibnamefont{and}
  \bibinfo{author}{\bibfnamefont{V.~P.} \bibnamefont{Krainov}},
  \emph{\bibinfo{title}{Above-Threshold Ionization of Atoms}}
  (\bibinfo{publisher}{Springer Berlin Heidelberg}, \bibinfo{address}{Berlin,
  Heidelberg}, \bibinfo{year}{2000}), pp. \bibinfo{pages}{151--187}.

\bibitem[{\citenamefont{Ivanov et~al.}(2005)\citenamefont{Ivanov, Spanner, and
  Smirnova}}]{Ivanov05}
\bibinfo{author}{\bibfnamefont{M.~Y.} \bibnamefont{Ivanov}},
  \bibinfo{author}{\bibfnamefont{M.}~\bibnamefont{Spanner}}, \bibnamefont{and}
  \bibinfo{author}{\bibfnamefont{O.}~\bibnamefont{Smirnova}},
  \bibinfo{journal}{J.\ Mod.\ Opt.} \textbf{\bibinfo{volume}{52}},
  \bibinfo{pages}{165} (\bibinfo{year}{2005}).

\bibitem[{\citenamefont{Keldysh}(1965)}]{Keldysh}
\bibinfo{author}{\bibfnamefont{L.}~\bibnamefont{Keldysh}},
  \bibinfo{journal}{JETP} \textbf{\bibinfo{volume}{20}}, \bibinfo{pages}{1307}
  (\bibinfo{year}{1965}).

\bibitem[{\citenamefont{Topcu and Robicheaux}(2012)}]{Topcu12}
\bibinfo{author}{\bibfnamefont{T.}~\bibnamefont{Topcu}} \bibnamefont{and}
  \bibinfo{author}{\bibfnamefont{F.}~\bibnamefont{Robicheaux}},
  \bibinfo{journal}{Phys. Rev. A} \textbf{\bibinfo{volume}{86}},
  \bibinfo{pages}{053407} (\bibinfo{year}{2012}).

\bibitem[{\citenamefont{Delone and Krainov}(1998)}]{Delone_1998}
\bibinfo{author}{\bibfnamefont{N.~B.} \bibnamefont{Delone}} \bibnamefont{and}
  \bibinfo{author}{\bibfnamefont{V.~P.} \bibnamefont{Krainov}},
  \bibinfo{journal}{Phys.\ Usp.} \textbf{\bibinfo{volume}{41}},
  \bibinfo{pages}{469} (\bibinfo{year}{1998}).

\bibitem[{\citenamefont{Reiss}(2010)}]{Reiss2010}
\bibinfo{author}{\bibfnamefont{H.~R.} \bibnamefont{Reiss}},
  \bibinfo{journal}{Phys. Rev. A} \textbf{\bibinfo{volume}{82}},
  \bibinfo{pages}{023418} (\bibinfo{year}{2010}).

\bibitem[{\citenamefont{Trombetta et~al.}(1989)\citenamefont{Trombetta, Basile,
  and Ferrante}}]{Trombetta89}
\bibinfo{author}{\bibfnamefont{F.}~\bibnamefont{Trombetta}},
  \bibinfo{author}{\bibfnamefont{S.}~\bibnamefont{Basile}}, \bibnamefont{and}
  \bibinfo{author}{\bibfnamefont{G.}~\bibnamefont{Ferrante}},
  \bibinfo{journal}{Phys. Rev. A} \textbf{\bibinfo{volume}{40}},
  \bibinfo{pages}{2774} (\bibinfo{year}{1989}).

\bibitem[{\citenamefont{Mishima et~al.}(2002)\citenamefont{Mishima, Hayashi,
  Yi, Lin, Selzle, and Schlag}}]{Mishima02}
\bibinfo{author}{\bibfnamefont{K.}~\bibnamefont{Mishima}},
  \bibinfo{author}{\bibfnamefont{M.}~\bibnamefont{Hayashi}},
  \bibinfo{author}{\bibfnamefont{J.}~\bibnamefont{Yi}},
  \bibinfo{author}{\bibfnamefont{S.~H.} \bibnamefont{Lin}},
  \bibinfo{author}{\bibfnamefont{H.~L.} \bibnamefont{Selzle}},
  \bibnamefont{and} \bibinfo{author}{\bibfnamefont{E.~W.}
  \bibnamefont{Schlag}}, \bibinfo{journal}{Phys. Rev. A}
  \textbf{\bibinfo{volume}{66}}, \bibinfo{pages}{033401}
  (\bibinfo{year}{2002}).

\bibitem[{\citenamefont{Chetty et~al.}(2022)\citenamefont{Chetty, Glover, Tong,
  deHarak, Xu, Haram, Bartschat, Palmer, Luiten, Light et~al.}}]{Chetty22}
\bibinfo{author}{\bibfnamefont{D.}~\bibnamefont{Chetty}},
  \bibinfo{author}{\bibfnamefont{R.~D.} \bibnamefont{Glover}},
  \bibinfo{author}{\bibfnamefont{X.~M.} \bibnamefont{Tong}},
  \bibinfo{author}{\bibfnamefont{B.~A.} \bibnamefont{deHarak}},
  \bibinfo{author}{\bibfnamefont{H.}~\bibnamefont{Xu}},
  \bibinfo{author}{\bibfnamefont{N.}~\bibnamefont{Haram}},
  \bibinfo{author}{\bibfnamefont{K.}~\bibnamefont{Bartschat}},
  \bibinfo{author}{\bibfnamefont{A.~J.} \bibnamefont{Palmer}},
  \bibinfo{author}{\bibfnamefont{A.~N.} \bibnamefont{Luiten}},
  \bibinfo{author}{\bibfnamefont{P.~S.} \bibnamefont{Light}},
  \bibnamefont{et~al.}, \bibinfo{journal}{Phys. Rev. Lett.}
  \textbf{\bibinfo{volume}{128}}, \bibinfo{pages}{173201}
  (\bibinfo{year}{2022}).

\bibitem[{\citenamefont{Karamatskou et~al.}(2013)\citenamefont{Karamatskou,
  Pabst, and Santra}}]{Karamatskou13}
\bibinfo{author}{\bibfnamefont{A.}~\bibnamefont{Karamatskou}},
  \bibinfo{author}{\bibfnamefont{S.}~\bibnamefont{Pabst}}, \bibnamefont{and}
  \bibinfo{author}{\bibfnamefont{R.}~\bibnamefont{Santra}},
  \bibinfo{journal}{Phys. Rev. A} \textbf{\bibinfo{volume}{87}},
  \bibinfo{pages}{043422} (\bibinfo{year}{2013}).

\bibitem[{\citenamefont{Douguet and Bartschat}(2018)}]{Douguet18}
\bibinfo{author}{\bibfnamefont{N.}~\bibnamefont{Douguet}} \bibnamefont{and}
  \bibinfo{author}{\bibfnamefont{K.}~\bibnamefont{Bartschat}},
  \bibinfo{journal}{Phys. Rev. A} \textbf{\bibinfo{volume}{97}},
  \bibinfo{pages}{013402} (\bibinfo{year}{2018}).

\bibitem[{\citenamefont{Krainov}(1995)}]{KRAINOV95}
\bibinfo{author}{\bibfnamefont{V.~P.} \bibnamefont{Krainov}},
  \bibinfo{journal}{J.\ Nonlin.\ Opt.\ Phys. \& Mat.}
  \textbf{\bibinfo{volume}{04}}, \bibinfo{pages}{775} (\bibinfo{year}{1995}).

\bibitem[{\citenamefont{Schuricke et~al.}(2011)\citenamefont{Schuricke, Zhu,
  Steinmann, Simeonidis, Ivanov, Kheifets, Grum-Grzhimailo, Bartschat, Dorn,
  and Ullrich}}]{Schuricke11}
\bibinfo{author}{\bibfnamefont{M.}~\bibnamefont{Schuricke}},
  \bibinfo{author}{\bibfnamefont{G.}~\bibnamefont{Zhu}},
  \bibinfo{author}{\bibfnamefont{J.}~\bibnamefont{Steinmann}},
  \bibinfo{author}{\bibfnamefont{K.}~\bibnamefont{Simeonidis}},
  \bibinfo{author}{\bibfnamefont{I.}~\bibnamefont{Ivanov}},
  \bibinfo{author}{\bibfnamefont{A.}~\bibnamefont{Kheifets}},
  \bibinfo{author}{\bibfnamefont{A.~N.} \bibnamefont{Grum-Grzhimailo}},
  \bibinfo{author}{\bibfnamefont{K.}~\bibnamefont{Bartschat}},
  \bibinfo{author}{\bibfnamefont{A.}~\bibnamefont{Dorn}}, \bibnamefont{and}
  \bibinfo{author}{\bibfnamefont{J.}~\bibnamefont{Ullrich}},
  \bibinfo{journal}{Phys. Rev. A} \textbf{\bibinfo{volume}{83}},
  \bibinfo{pages}{023413} (\bibinfo{year}{2011}).

\bibitem[{\citenamefont{Arb\'o et~al.}(2008)\citenamefont{Arb\'o, Miraglia,
  Gravielle, Schiessl, Persson, and Burgd\"orfer}}]{Arbo08}
\bibinfo{author}{\bibfnamefont{D.~G.} \bibnamefont{Arb\'o}},
  \bibinfo{author}{\bibfnamefont{J.~E.} \bibnamefont{Miraglia}},
  \bibinfo{author}{\bibfnamefont{M.~S.} \bibnamefont{Gravielle}},
  \bibinfo{author}{\bibfnamefont{K.}~\bibnamefont{Schiessl}},
  \bibinfo{author}{\bibfnamefont{E.}~\bibnamefont{Persson}}, \bibnamefont{and}
  \bibinfo{author}{\bibfnamefont{J.}~\bibnamefont{Burgd\"orfer}},
  \bibinfo{journal}{Phys. Rev. A} \textbf{\bibinfo{volume}{77}},
  \bibinfo{pages}{013401} (\bibinfo{year}{2008}).

\bibitem[{\citenamefont{Popruzhenko et~al.}(2008)\citenamefont{Popruzhenko,
  Mur, Popov, and Bauer}}]{Popruzhenko08}
\bibinfo{author}{\bibfnamefont{S.~V.} \bibnamefont{Popruzhenko}},
  \bibinfo{author}{\bibfnamefont{V.~D.} \bibnamefont{Mur}},
  \bibinfo{author}{\bibfnamefont{V.~S.} \bibnamefont{Popov}}, \bibnamefont{and}
  \bibinfo{author}{\bibfnamefont{D.}~\bibnamefont{Bauer}},
  \bibinfo{journal}{Phys. Rev. Lett.} \textbf{\bibinfo{volume}{101}},
  \bibinfo{pages}{193003} (\bibinfo{year}{2008}).

\bibitem[{\citenamefont{Wang et~al.}(2019)\citenamefont{Wang, Zhang, Li, Xu,
  Cao, Zhou, Cao, and Lu}}]{Wang19}
\bibinfo{author}{\bibfnamefont{R.}~\bibnamefont{Wang}},
  \bibinfo{author}{\bibfnamefont{Q.}~\bibnamefont{Zhang}},
  \bibinfo{author}{\bibfnamefont{D.}~\bibnamefont{Li}},
  \bibinfo{author}{\bibfnamefont{S.}~\bibnamefont{Xu}},
  \bibinfo{author}{\bibfnamefont{P.}~\bibnamefont{Cao}},
  \bibinfo{author}{\bibfnamefont{Y.}~\bibnamefont{Zhou}},
  \bibinfo{author}{\bibfnamefont{W.}~\bibnamefont{Cao}}, \bibnamefont{and}
  \bibinfo{author}{\bibfnamefont{P.}~\bibnamefont{Lu}}, \bibinfo{journal}{Opt.
  Express} \textbf{\bibinfo{volume}{27}}, \bibinfo{pages}{6471}
  (\bibinfo{year}{2019}).

\bibitem[{\citenamefont{Zheltikov}(2016)}]{Zheltikov16}
\bibinfo{author}{\bibfnamefont{A.~M.} \bibnamefont{Zheltikov}},
  \bibinfo{journal}{Phys. Rev. A} \textbf{\bibinfo{volume}{94}},
  \bibinfo{pages}{043412} (\bibinfo{year}{2016}).

\bibitem[{\citenamefont{Heldt et~al.}(2023)\citenamefont{Heldt, Dubois, Birk,
  Borisova, Lando, Ott, and Pfeifer}}]{Heldt23}
\bibinfo{author}{\bibfnamefont{T.}~\bibnamefont{Heldt}},
  \bibinfo{author}{\bibfnamefont{J.}~\bibnamefont{Dubois}},
  \bibinfo{author}{\bibfnamefont{P.}~\bibnamefont{Birk}},
  \bibinfo{author}{\bibfnamefont{G.~D.} \bibnamefont{Borisova}},
  \bibinfo{author}{\bibfnamefont{G.~M.} \bibnamefont{Lando}},
  \bibinfo{author}{\bibfnamefont{C.}~\bibnamefont{Ott}}, \bibnamefont{and}
  \bibinfo{author}{\bibfnamefont{T.}~\bibnamefont{Pfeifer}},
  \bibinfo{journal}{Phys. Rev. Lett.} \textbf{\bibinfo{volume}{130}},
  \bibinfo{pages}{183201} (\bibinfo{year}{2023}).

\bibitem[{\citenamefont{Xie et~al.}(2022)\citenamefont{Xie, Li, Zhou, and
  Lu}}]{Xie22}
\bibinfo{author}{\bibfnamefont{W.}~\bibnamefont{Xie}},
  \bibinfo{author}{\bibfnamefont{M.}~\bibnamefont{Li}},
  \bibinfo{author}{\bibfnamefont{Y.}~\bibnamefont{Zhou}}, \bibnamefont{and}
  \bibinfo{author}{\bibfnamefont{P.}~\bibnamefont{Lu}}, \bibinfo{journal}{Phys.
  Rev. A} \textbf{\bibinfo{volume}{105}}, \bibinfo{pages}{013119}
  (\bibinfo{year}{2022}).

\bibitem[{\citenamefont{Ni et~al.}(2016)\citenamefont{Ni, Saalmann, and
  Rost}}]{Ni16}
\bibinfo{author}{\bibfnamefont{H.}~\bibnamefont{Ni}},
  \bibinfo{author}{\bibfnamefont{U.}~\bibnamefont{Saalmann}}, \bibnamefont{and}
  \bibinfo{author}{\bibfnamefont{J.-M.} \bibnamefont{Rost}},
  \bibinfo{journal}{Phys. Rev. Lett.} \textbf{\bibinfo{volume}{117}},
  \bibinfo{pages}{023002} (\bibinfo{year}{2016}).

\bibitem[{\citenamefont{Camus et~al.}(2017)\citenamefont{Camus, Yakaboylu,
  Fechner, Klaiber, Laux, Mi, Hatsagortsyan, Pfeifer, Keitel, and
  Moshammer}}]{Camus17}
\bibinfo{author}{\bibfnamefont{N.}~\bibnamefont{Camus}},
  \bibinfo{author}{\bibfnamefont{E.}~\bibnamefont{Yakaboylu}},
  \bibinfo{author}{\bibfnamefont{L.}~\bibnamefont{Fechner}},
  \bibinfo{author}{\bibfnamefont{M.}~\bibnamefont{Klaiber}},
  \bibinfo{author}{\bibfnamefont{M.}~\bibnamefont{Laux}},
  \bibinfo{author}{\bibfnamefont{Y.}~\bibnamefont{Mi}},
  \bibinfo{author}{\bibfnamefont{K.~Z.} \bibnamefont{Hatsagortsyan}},
  \bibinfo{author}{\bibfnamefont{T.}~\bibnamefont{Pfeifer}},
  \bibinfo{author}{\bibfnamefont{C.~H.} \bibnamefont{Keitel}},
  \bibnamefont{and}
  \bibinfo{author}{\bibfnamefont{R.}~\bibnamefont{Moshammer}},
  \bibinfo{journal}{Phys. Rev. Lett.} \textbf{\bibinfo{volume}{119}},
  \bibinfo{pages}{023201} (\bibinfo{year}{2017}).

\bibitem[{\citenamefont{Salières et~al.}(2001)\citenamefont{Salières, Carré,
  Déroff, Grasbon, Paulus, Walther, Kopold, Becker, Milošević, Sanpera
  et~al.}}]{Salieres01}
\bibinfo{author}{\bibfnamefont{P.}~\bibnamefont{Salières}},
  \bibinfo{author}{\bibfnamefont{B.}~\bibnamefont{Carré}},
  \bibinfo{author}{\bibfnamefont{L.~L.} \bibnamefont{Déroff}},
  \bibinfo{author}{\bibfnamefont{F.}~\bibnamefont{Grasbon}},
  \bibinfo{author}{\bibfnamefont{G.~G.} \bibnamefont{Paulus}},
  \bibinfo{author}{\bibfnamefont{H.}~\bibnamefont{Walther}},
  \bibinfo{author}{\bibfnamefont{R.}~\bibnamefont{Kopold}},
  \bibinfo{author}{\bibfnamefont{W.}~\bibnamefont{Becker}},
  \bibinfo{author}{\bibfnamefont{D.~B.} \bibnamefont{Milošević}},
  \bibinfo{author}{\bibfnamefont{A.}~\bibnamefont{Sanpera}},
  \bibnamefont{et~al.}, \bibinfo{journal}{Science}
  \textbf{\bibinfo{volume}{292}}, \bibinfo{pages}{902} (\bibinfo{year}{2001}).

\bibitem[{\citenamefont{Yang and Robicheaux}(2016)}]{Yang16}
\bibinfo{author}{\bibfnamefont{B.~C.} \bibnamefont{Yang}} \bibnamefont{and}
  \bibinfo{author}{\bibfnamefont{F.}~\bibnamefont{Robicheaux}},
  \bibinfo{journal}{Phys. Rev. A} \textbf{\bibinfo{volume}{93}},
  \bibinfo{pages}{053413} (\bibinfo{year}{2016}).

\bibitem[{\citenamefont{Tan et~al.}(2021)\citenamefont{Tan, Zhou, Xu, Ke,
  Liang, Ma, Cao, Li, Zhang, and Lu}}]{Tan21}
\bibinfo{author}{\bibfnamefont{J.}~\bibnamefont{Tan}},
  \bibinfo{author}{\bibfnamefont{Y.}~\bibnamefont{Zhou}},
  \bibinfo{author}{\bibfnamefont{S.}~\bibnamefont{Xu}},
  \bibinfo{author}{\bibfnamefont{Q.}~\bibnamefont{Ke}},
  \bibinfo{author}{\bibfnamefont{J.}~\bibnamefont{Liang}},
  \bibinfo{author}{\bibfnamefont{X.}~\bibnamefont{Ma}},
  \bibinfo{author}{\bibfnamefont{W.}~\bibnamefont{Cao}},
  \bibinfo{author}{\bibfnamefont{M.}~\bibnamefont{Li}},
  \bibinfo{author}{\bibfnamefont{Q.}~\bibnamefont{Zhang}}, \bibnamefont{and}
  \bibinfo{author}{\bibfnamefont{P.}~\bibnamefont{Lu}}, \bibinfo{journal}{Opt.\
  Expr.} \textbf{\bibinfo{volume}{29}}, \bibinfo{pages}{37927}
  (\bibinfo{year}{2021}).

\bibitem[{\citenamefont{Krause et~al.}(1992)\citenamefont{Krause, Schafer, and
  Kulander}}]{PhysRevLett.68.3535}
\bibinfo{author}{\bibfnamefont{J.~L.} \bibnamefont{Krause}},
  \bibinfo{author}{\bibfnamefont{K.~J.} \bibnamefont{Schafer}},
  \bibnamefont{and} \bibinfo{author}{\bibfnamefont{K.~C.}
  \bibnamefont{Kulander}}, \bibinfo{journal}{Phys. Rev. Lett.}
  \textbf{\bibinfo{volume}{68}}, \bibinfo{pages}{3535} (\bibinfo{year}{1992}).

\bibitem[{\citenamefont{Corkum}(1993)}]{corkum1993}
\bibinfo{author}{\bibfnamefont{P.~B.} \bibnamefont{Corkum}},
  \bibinfo{journal}{Phys. Rev. Lett.} \textbf{\bibinfo{volume}{71}},
  \bibinfo{pages}{1994} (\bibinfo{year}{1993}).

\bibitem[{\citenamefont{Corkum and Krausz}(2007)}]{Corkum07}
\bibinfo{author}{\bibfnamefont{P.~B.} \bibnamefont{Corkum}} \bibnamefont{and}
  \bibinfo{author}{\bibfnamefont{F.}~\bibnamefont{Krausz}},
  \bibinfo{journal}{Nat. Phys.} \textbf{\bibinfo{volume}{3}},
  \bibinfo{pages}{381} (\bibinfo{year}{2007}).

\bibitem[{\citenamefont{{Perelomov} et~al.}(1966)\citenamefont{{Perelomov},
  {Popov}, and {Terent'ev}}}]{Perelomov66}
\bibinfo{author}{\bibfnamefont{A.~M.} \bibnamefont{{Perelomov}}},
  \bibinfo{author}{\bibfnamefont{V.~S.} \bibnamefont{{Popov}}},
  \bibnamefont{and} \bibinfo{author}{\bibfnamefont{M.~V.}
  \bibnamefont{{Terent'ev}}}, \bibinfo{journal}{Sov.\ J. Expt. Theor. Phy.
  (JETP)} \textbf{\bibinfo{volume}{23}}, \bibinfo{pages}{924}
  (\bibinfo{year}{1966}).

\bibitem[{\citenamefont{Hack et~al.}(2021)\citenamefont{Hack, Majorosi,
  Benedict, Varr\'o, and Czirj\'ak}}]{Hack21}
\bibinfo{author}{\bibfnamefont{S.}~\bibnamefont{Hack}},
  \bibinfo{author}{\bibfnamefont{S.}~\bibnamefont{Majorosi}},
  \bibinfo{author}{\bibfnamefont{M.~G.} \bibnamefont{Benedict}},
  \bibinfo{author}{\bibfnamefont{S.}~\bibnamefont{Varr\'o}}, \bibnamefont{and}
  \bibinfo{author}{\bibfnamefont{A.}~\bibnamefont{Czirj\'ak}},
  \bibinfo{journal}{Phys. Rev. A} \textbf{\bibinfo{volume}{104}},
  \bibinfo{pages}{L031102} (\bibinfo{year}{2021}).

\bibitem[{\citenamefont{D\"{u}rr and Teufel}(2009)}]{duerr2009bohmian}
\bibinfo{author}{\bibfnamefont{D.}~\bibnamefont{D\"{u}rr}} \bibnamefont{and}
  \bibinfo{author}{\bibfnamefont{S.}~\bibnamefont{Teufel}},
  \emph{\bibinfo{title}{Bohmian Mechanics}} (\bibinfo{publisher}{Springer
  Berlin Heidelberg}, \bibinfo{year}{2009}).

\bibitem[{\citenamefont{Botheron and Pons}(2010)}]{Botheron10}
\bibinfo{author}{\bibfnamefont{P.}~\bibnamefont{Botheron}} \bibnamefont{and}
  \bibinfo{author}{\bibfnamefont{B.}~\bibnamefont{Pons}},
  \bibinfo{journal}{Phys. Rev. A} \textbf{\bibinfo{volume}{82}},
  \bibinfo{pages}{021404} (\bibinfo{year}{2010}).

\bibitem[{\citenamefont{Benseny et~al.}(2014)\citenamefont{Benseny, Albareda,
  Sanz, Mompart, and Oriols}}]{Benseny14}
\bibinfo{author}{\bibfnamefont{A.}~\bibnamefont{Benseny}},
  \bibinfo{author}{\bibfnamefont{G.}~\bibnamefont{Albareda}},
  \bibinfo{author}{\bibfnamefont{A.~S.} \bibnamefont{Sanz}},
  \bibinfo{author}{\bibfnamefont{J.}~\bibnamefont{Mompart}}, \bibnamefont{and}
  \bibinfo{author}{\bibfnamefont{X.}~\bibnamefont{Oriols}},
  \bibinfo{journal}{Eur. J. Phys. D} \textbf{\bibinfo{volume}{68}},
  \bibinfo{pages}{286} (\bibinfo{year}{2014}).

\bibitem[{\citenamefont{Jooya et~al.}(2015)\citenamefont{Jooya, Telnov, Li, and
  Chu}}]{Jooya15}
\bibinfo{author}{\bibfnamefont{H.~Z.} \bibnamefont{Jooya}},
  \bibinfo{author}{\bibfnamefont{D.~A.} \bibnamefont{Telnov}},
  \bibinfo{author}{\bibfnamefont{P.-C.} \bibnamefont{Li}}, \bibnamefont{and}
  \bibinfo{author}{\bibfnamefont{S.-I.} \bibnamefont{Chu}},
  \bibinfo{journal}{Phys. Rev. A} \textbf{\bibinfo{volume}{91}},
  \bibinfo{pages}{063412} (\bibinfo{year}{2015}).

\bibitem[{\citenamefont{Zimmermann et~al.}(2016)\citenamefont{Zimmermann,
  Mishra, Doran, Gordon, and Landsman}}]{Zimmermann16}
\bibinfo{author}{\bibfnamefont{T.}~\bibnamefont{Zimmermann}},
  \bibinfo{author}{\bibfnamefont{S.}~\bibnamefont{Mishra}},
  \bibinfo{author}{\bibfnamefont{B.~R.} \bibnamefont{Doran}},
  \bibinfo{author}{\bibfnamefont{D.~F.} \bibnamefont{Gordon}},
  \bibnamefont{and} \bibinfo{author}{\bibfnamefont{A.~S.}
  \bibnamefont{Landsman}}, \bibinfo{journal}{Phys. Rev. Lett.}
  \textbf{\bibinfo{volume}{116}}, \bibinfo{pages}{233603}
  (\bibinfo{year}{2016}).

\bibitem[{\citenamefont{Douguet et~al.}(2016)\citenamefont{Douguet,
  Grum-Grzhimailo, Gryzlova, Staroselskaya, Venzke, and Bartschat}}]{Douguet16}
\bibinfo{author}{\bibfnamefont{N.}~\bibnamefont{Douguet}},
  \bibinfo{author}{\bibfnamefont{A.~N.} \bibnamefont{Grum-Grzhimailo}},
  \bibinfo{author}{\bibfnamefont{E.~V.} \bibnamefont{Gryzlova}},
  \bibinfo{author}{\bibfnamefont{E.~I.} \bibnamefont{Staroselskaya}},
  \bibinfo{author}{\bibfnamefont{J.}~\bibnamefont{Venzke}}, \bibnamefont{and}
  \bibinfo{author}{\bibfnamefont{K.}~\bibnamefont{Bartschat}},
  \bibinfo{journal}{Phys. Rev. A} \textbf{\bibinfo{volume}{93}},
  \bibinfo{pages}{033402} (\bibinfo{year}{2016}).

\bibitem[{\citenamefont{Song et~al.}(2017)\citenamefont{Song, Yang, Guo, and
  Li}}]{Song_2017}
\bibinfo{author}{\bibfnamefont{Y.}~\bibnamefont{Song}},
  \bibinfo{author}{\bibfnamefont{Y.}~\bibnamefont{Yang}},
  \bibinfo{author}{\bibfnamefont{F.}~\bibnamefont{Guo}}, \bibnamefont{and}
  \bibinfo{author}{\bibfnamefont{S.}~\bibnamefont{Li}}, \bibinfo{journal}{J.\
  Phys.\ B: At.\ Mol.\ Opt.\ Phys.} \textbf{\bibinfo{volume}{50}},
  \bibinfo{pages}{095003} (\bibinfo{year}{2017}).

\bibitem[{\citenamefont{Kobe and Yang}(1987)}]{Kobe_1987}
\bibinfo{author}{\bibfnamefont{D.~H.} \bibnamefont{Kobe}} \bibnamefont{and}
  \bibinfo{author}{\bibfnamefont{K.-H.} \bibnamefont{Yang}},
  \bibinfo{journal}{Eur.\ J.\ Phys.} \textbf{\bibinfo{volume}{8}},
  \bibinfo{pages}{236} (\bibinfo{year}{1987}).

\bibitem[{\citenamefont{Kim et~al.}(2021)\citenamefont{Kim, Schmude, Burkard,
  and Moskalenko}}]{Kim_2021}
\bibinfo{author}{\bibfnamefont{S.}~\bibnamefont{Kim}},
  \bibinfo{author}{\bibfnamefont{T.}~\bibnamefont{Schmude}},
  \bibinfo{author}{\bibfnamefont{G.}~\bibnamefont{Burkard}}, \bibnamefont{and}
  \bibinfo{author}{\bibfnamefont{A.~S.} \bibnamefont{Moskalenko}},
  \bibinfo{journal}{New J.~Phys.} \textbf{\bibinfo{volume}{23}},
  \bibinfo{pages}{083006} (\bibinfo{year}{2021}).

\bibitem[{\citenamefont{Ivanov et~al.}(2024)\citenamefont{Ivanov, Kheifets,
  Schimmoller, Landsman, and Kim}}]{Ivanov24}
\bibinfo{author}{\bibfnamefont{I.~A.} \bibnamefont{Ivanov}},
  \bibinfo{author}{\bibfnamefont{A.~S.} \bibnamefont{Kheifets}},
  \bibinfo{author}{\bibfnamefont{A.}~\bibnamefont{Schimmoller}},
  \bibinfo{author}{\bibfnamefont{A.~S.} \bibnamefont{Landsman}},
  \bibnamefont{and} \bibinfo{author}{\bibfnamefont{K.~T.} \bibnamefont{Kim}},
  \bibinfo{journal}{Phys. Rev. Res.} \textbf{\bibinfo{volume}{6}},
  \bibinfo{pages}{023049} (\bibinfo{year}{2024}).

\bibitem[{\citenamefont{Han et~al.}(2019)\citenamefont{Han, Ge, Fang, Yu, Guo,
  Ma, Deng, Gong, and Liu}}]{Han19}
\bibinfo{author}{\bibfnamefont{M.}~\bibnamefont{Han}},
  \bibinfo{author}{\bibfnamefont{P.}~\bibnamefont{Ge}},
  \bibinfo{author}{\bibfnamefont{Y.}~\bibnamefont{Fang}},
  \bibinfo{author}{\bibfnamefont{X.}~\bibnamefont{Yu}},
  \bibinfo{author}{\bibfnamefont{Z.}~\bibnamefont{Guo}},
  \bibinfo{author}{\bibfnamefont{X.}~\bibnamefont{Ma}},
  \bibinfo{author}{\bibfnamefont{Y.}~\bibnamefont{Deng}},
  \bibinfo{author}{\bibfnamefont{Q.}~\bibnamefont{Gong}}, \bibnamefont{and}
  \bibinfo{author}{\bibfnamefont{Y.}~\bibnamefont{Liu}},
  \bibinfo{journal}{Phys. Rev. Lett.} \textbf{\bibinfo{volume}{123}},
  \bibinfo{pages}{073201} (\bibinfo{year}{2019}).

\bibitem[{\citenamefont{Garashchuk and Rassolov}(2003)}]{Garashchuk03}
\bibinfo{author}{\bibfnamefont{S.}~\bibnamefont{Garashchuk}} \bibnamefont{and}
  \bibinfo{author}{\bibfnamefont{V.~A.} \bibnamefont{Rassolov}},
  \bibinfo{journal}{J.\ Chem. Phys.} \textbf{\bibinfo{volume}{118}},
  \bibinfo{pages}{2482} (\bibinfo{year}{2003}).

\bibitem[{\citenamefont{Torlina et~al.}(2015)\citenamefont{Torlina, Morales,
  Kaushal, Ivanov, Kheifets, Zielinski, Scrinzi, Muller, Sukiasyan, Ivanov
  et~al.}}]{Torlina15}
\bibinfo{author}{\bibfnamefont{L.}~\bibnamefont{Torlina}},
  \bibinfo{author}{\bibfnamefont{F.}~\bibnamefont{Morales}},
  \bibinfo{author}{\bibfnamefont{J.}~\bibnamefont{Kaushal}},
  \bibinfo{author}{\bibfnamefont{I.}~\bibnamefont{Ivanov}},
  \bibinfo{author}{\bibfnamefont{A.}~\bibnamefont{Kheifets}},
  \bibinfo{author}{\bibfnamefont{A.}~\bibnamefont{Zielinski}},
  \bibinfo{author}{\bibfnamefont{A.}~\bibnamefont{Scrinzi}},
  \bibinfo{author}{\bibfnamefont{H.~G.} \bibnamefont{Muller}},
  \bibinfo{author}{\bibfnamefont{S.}~\bibnamefont{Sukiasyan}},
  \bibinfo{author}{\bibfnamefont{M.}~\bibnamefont{Ivanov}},
  \bibnamefont{et~al.}, \bibinfo{journal}{Nat. Phys.}
  \textbf{\bibinfo{volume}{11}}, \bibinfo{pages}{503} (\bibinfo{year}{2015}).

\bibitem[{\citenamefont{Klaiber et~al.}(2015)\citenamefont{Klaiber,
  Hatsagortsyan, and Keitel}}]{Klaiber15}
\bibinfo{author}{\bibfnamefont{M.}~\bibnamefont{Klaiber}},
  \bibinfo{author}{\bibfnamefont{K.~Z.} \bibnamefont{Hatsagortsyan}},
  \bibnamefont{and} \bibinfo{author}{\bibfnamefont{C.~H.}
  \bibnamefont{Keitel}}, \bibinfo{journal}{Phys. Rev. Lett.}
  \textbf{\bibinfo{volume}{114}}, \bibinfo{pages}{083001}
  (\bibinfo{year}{2015}).

\bibitem[{\citenamefont{Ammosov et~al.}(1986)\citenamefont{Ammosov, Delone, and
  Krainov}}]{Ammosov86}
\bibinfo{author}{\bibfnamefont{M.~V.} \bibnamefont{Ammosov}},
  \bibinfo{author}{\bibfnamefont{N.~B.} \bibnamefont{Delone}},
  \bibnamefont{and} \bibinfo{author}{\bibfnamefont{V.~P.}
  \bibnamefont{Krainov}}, in \emph{\bibinfo{booktitle}{High Intensity Laser
  Processes}}, edited by \bibinfo{editor}{\bibfnamefont{J.~A.}
  \bibnamefont{Alcock}}, \bibinfo{organization}{International Society for
  Optics and Photonics} (\bibinfo{publisher}{SPIE}, \bibinfo{year}{1986}), vol.
  \bibinfo{volume}{0664}, pp. \bibinfo{pages}{138 -- 141}.

\bibitem[{\citenamefont{Zimmermann et~al.}(2017)\citenamefont{Zimmermann,
  Patchkovskii, Ivanov, and Eichmann}}]{Zimmermann17}
\bibinfo{author}{\bibfnamefont{H.}~\bibnamefont{Zimmermann}},
  \bibinfo{author}{\bibfnamefont{S.}~\bibnamefont{Patchkovskii}},
  \bibinfo{author}{\bibfnamefont{M.}~\bibnamefont{Ivanov}}, \bibnamefont{and}
  \bibinfo{author}{\bibfnamefont{U.}~\bibnamefont{Eichmann}},
  \bibinfo{journal}{Phys. Rev. Lett.} \textbf{\bibinfo{volume}{118}},
  \bibinfo{pages}{013003} (\bibinfo{year}{2017}).

\bibitem[{\citenamefont{Ishii et~al.}(2008)\citenamefont{Ishii, Kosuge,
  Hayashi, Kanai, Itatani, Adachi, and Watanabe}}]{Ishii08}
\bibinfo{author}{\bibfnamefont{N.}~\bibnamefont{Ishii}},
  \bibinfo{author}{\bibfnamefont{A.}~\bibnamefont{Kosuge}},
  \bibinfo{author}{\bibfnamefont{T.}~\bibnamefont{Hayashi}},
  \bibinfo{author}{\bibfnamefont{T.}~\bibnamefont{Kanai}},
  \bibinfo{author}{\bibfnamefont{J.}~\bibnamefont{Itatani}},
  \bibinfo{author}{\bibfnamefont{S.}~\bibnamefont{Adachi}}, \bibnamefont{and}
  \bibinfo{author}{\bibfnamefont{S.}~\bibnamefont{Watanabe}},
  \bibinfo{journal}{Opt. Express} \textbf{\bibinfo{volume}{16}},
  \bibinfo{pages}{20876} (\bibinfo{year}{2008}).

\bibitem[{\citenamefont{Dong et~al.}(2020)\citenamefont{Dong, Hu, and
  Zhao}}]{Dong20}
\bibinfo{author}{\bibfnamefont{W.}~\bibnamefont{Dong}},
  \bibinfo{author}{\bibfnamefont{H.}~\bibnamefont{Hu}}, \bibnamefont{and}
  \bibinfo{author}{\bibfnamefont{Z.}~\bibnamefont{Zhao}},
  \bibinfo{journal}{Opt. Express} \textbf{\bibinfo{volume}{28}},
  \bibinfo{pages}{22490} (\bibinfo{year}{2020}).

\bibitem[{\citenamefont{Petrakis et~al.}(2021)\citenamefont{Petrakis,
  Bakarezos, Tatarakis, Benis, and Papadogiannis}}]{Petrakis21}
\bibinfo{author}{\bibfnamefont{S.}~\bibnamefont{Petrakis}},
  \bibinfo{author}{\bibfnamefont{M.}~\bibnamefont{Bakarezos}},
  \bibinfo{author}{\bibfnamefont{M.}~\bibnamefont{Tatarakis}},
  \bibinfo{author}{\bibfnamefont{E.~P.} \bibnamefont{Benis}}, \bibnamefont{and}
  \bibinfo{author}{\bibfnamefont{M.~A.} \bibnamefont{Papadogiannis}},
  \bibinfo{journal}{Sci. Rep.} \textbf{\bibinfo{volume}{11}},
  \bibinfo{pages}{23882} (\bibinfo{year}{2021}).

\bibitem[{\citenamefont{Piper}(2022)}]{Piper22}
\bibinfo{author}{\bibfnamefont{A.}~\bibnamefont{Piper}},
  \emph{\bibinfo{title}{The Strong Field Simulator: Studying Quantum
  Trajectories in Classical Fields}} (\bibinfo{publisher}{ProQuest
  Dissertations \& Theses Global}, \bibinfo{year}{2022}).

\bibitem[{\citenamefont{Giovannini et~al.}(2023)\citenamefont{Giovannini,
  Küpper, and Trabattoni}}]{DeGiovannini_2023}
\bibinfo{author}{\bibfnamefont{U.~D.} \bibnamefont{Giovannini}},
  \bibinfo{author}{\bibfnamefont{J.}~\bibnamefont{Küpper}}, \bibnamefont{and}
  \bibinfo{author}{\bibfnamefont{A.}~\bibnamefont{Trabattoni}},
  \bibinfo{journal}{Journal of Physics B: Atomic, Molecular and Optical
  Physics} \textbf{\bibinfo{volume}{56}}, \bibinfo{pages}{054002}
  (\bibinfo{year}{2023}).

\bibitem[{\citenamefont{Nubbemeyer et~al.}(2008)\citenamefont{Nubbemeyer,
  Gorling, Saenz, Eichmann, and Sandner}}]{Nubbemeyer08}
\bibinfo{author}{\bibfnamefont{T.}~\bibnamefont{Nubbemeyer}},
  \bibinfo{author}{\bibfnamefont{K.}~\bibnamefont{Gorling}},
  \bibinfo{author}{\bibfnamefont{A.}~\bibnamefont{Saenz}},
  \bibinfo{author}{\bibfnamefont{U.}~\bibnamefont{Eichmann}}, \bibnamefont{and}
  \bibinfo{author}{\bibfnamefont{W.}~\bibnamefont{Sandner}},
  \bibinfo{journal}{Phys. Rev. Lett.} \textbf{\bibinfo{volume}{101}},
  \bibinfo{pages}{233001} (\bibinfo{year}{2008}).

\bibitem[{\citenamefont{Bloch et~al.}(2021)\citenamefont{Bloch, Larroque,
  Rozen, Beaulieu, Comby, Beauvarlet, Descamps, Fabre, Petit, Ta\"{\i}eb
  et~al.}}]{Bloch21}
\bibinfo{author}{\bibfnamefont{E.}~\bibnamefont{Bloch}},
  \bibinfo{author}{\bibfnamefont{S.}~\bibnamefont{Larroque}},
  \bibinfo{author}{\bibfnamefont{S.}~\bibnamefont{Rozen}},
  \bibinfo{author}{\bibfnamefont{S.}~\bibnamefont{Beaulieu}},
  \bibinfo{author}{\bibfnamefont{A.}~\bibnamefont{Comby}},
  \bibinfo{author}{\bibfnamefont{S.}~\bibnamefont{Beauvarlet}},
  \bibinfo{author}{\bibfnamefont{D.}~\bibnamefont{Descamps}},
  \bibinfo{author}{\bibfnamefont{B.}~\bibnamefont{Fabre}},
  \bibinfo{author}{\bibfnamefont{S.}~\bibnamefont{Petit}},
  \bibinfo{author}{\bibfnamefont{R.}~\bibnamefont{Ta\"{\i}eb}},
  \bibnamefont{et~al.}, \bibinfo{journal}{Phys. Rev. X}
  \textbf{\bibinfo{volume}{11}}, \bibinfo{pages}{041056}
  (\bibinfo{year}{2021}).

\end{thebibliography}
 
  
\end{document}